\documentclass{PoS}

\title{MeV Pulsars: Modeling Spectra and Polarization}

\ShortTitle{MeV Pulsars: Modeling Spectra and Polarization}

\author{\speaker{Alice K. Harding} \\
        NASA Goddard Space Flight Center\\
        E-mail: \email{Alice.K.Harding@nasa.gov}}

\author{Constantinos Kalapotharakos\\
        CRESST/University of Maryland\\
        NASA Goddard Space Flight Center\\
        E-mail: \email{konstantinos.kalapotharakos@nasa.gov}}

\abstract{
A sub-population of energetic rotation-powered pulsars show high fluxes of pulsed non-thermal hard X-ray emission.  While this `MeV pulsar' population includes some radio-loud pulsars like the Crab, a significant number have no detected radio or GeV emission, a mystery since gamma-ray emission is a common characteristic of pulsars with high spin-down power.  Their steeply rising hard X-ray spectral energy distributions (SEDs) suggest peaks at 0.1 - 1 MeV but they have not been detected above 200 keV.  Several upcoming and planned telescopes may shed light on the MeV pulsars.  The Neutron star Interior Composition ExploreR (NICER) will observe pulsars in the 0.2 - 12 keV band and may discover additional MeV pulsars.  Planned telescopes, such as All-Sky Medium-Energy Gamma-Ray Observatory (AMEGO) and e-ASTROGAM, will detect emission above 0.2 MeV and polarization in the 0.2 - 10 MeV band.  We present a model for the spectrum and polarization of MeV pulsars where the X-ray emission comes from electron-positron pairs radiating in the outer magnetosphere and current sheet.  This model predicts that the peak of the SED increases with surface magnetic field strength if the pairs are produced in polar cap cascades.  For small inclination angles, a range of viewing angles can miss both the radio pulse and the GeV pulse from particles accelerating near the current sheet.  Characterizing the emission and geometry of MeV pulsars can thus provide clues to the source of pairs and acceleration in the magnetosphere. }

\FullConference{7th International Fermi Symposium\\
        15-20 October 2017\\
        Garmisch-Partenkirchen, Germany}

\begin{document}

\section{Introduction}

The {\it Fermi} Large Area Telescope (LAT) has discovered over 200 $\gamma$-ray pulsars in the nearly ten years since its launch.  This population includes both young and very energetic pulsars, middle-aged pulsars and old, recycled millisecond pulsars \cite{Abdo2013}.  The vast majority of these have their spectral energy peaks (SEDs) at several GeV.  A few have bright GeV emission but SED peaks at hard X-ray energies, such as the Crab, and Crab-like pulsar PSR J0540, and a few have very soft Fermi LAT spectra and SED peaks at X-ray energies, such as PSR B1509-58.  There exists another population of very energetic, young rotation-powered pulsars having non-thermal hard X-ray emission with SEDs rising toward 1 MeV, but have no detected emission by Fermi LAT and are radio-quiet.  This group of around 11 sources \cite{KuiperHermsen2015} also have broad, single-peaked light curves in contrast to the narrow, double-peaked light curved of most GeV-bright pulsars.  

The pulsars with SED peaks at hard X-ray energies, which we call MeV pulsars, present the un-answered question: Why do a large sub-set of these with no detected GeV emission?  The unusual light curves of the GeV-quiet pulsars suggest that the answer may be geometry or viewing angle.  The non-thermal hard X-ray emission of these object also may provide a probe of the magnetospheric plasma production.  A variety of different emission models suggest that this emission results from synchrotron radiation of electron-positron pairs, produced either at the polar caps \cite{Harding2008} or in the outer gaps \cite{Takata2007}, with emission occurring in both cases in the outer magnetosphere near the light cylinder, $R_{\rm LC} = c/\Omega$, with $\Omega$ the pulsar rotation rate.  Since the pairs are thought to be the prime source of plasma for the pulsar magnetosphere, measuring the properties of the MeV pulsar spectra constrains the spectrum and density of the pairs.  Recent particle-in-cell (PIC) simulations of the global pulsar magnetosphere \cite{Philippov2014} show that the origin of the pair plasma determines the gap locations \cite{ChenBeloborodov2014}, the current composition \cite{Brambilla2017,PhilippovSpitkovsky2017} and the density of the high-energy particles \cite{Kala2017}. 

We have modeled the spectra and polarization properties of MeV pulsars assuming that the pairs that produce synchrotron radiation in the outer magnetosphere are produced in polar cap cascades.  We will show that the resulting shape of the spectrum reflects the pair spectrum, the flux level reflects the pair density and the energy of the SED peak can determine whether the pairs originate from the polar cap.  

\section{Spectral Models of MeV Pulsars}

\begin{figure}
\center{\includegraphics[width=.8\textwidth]{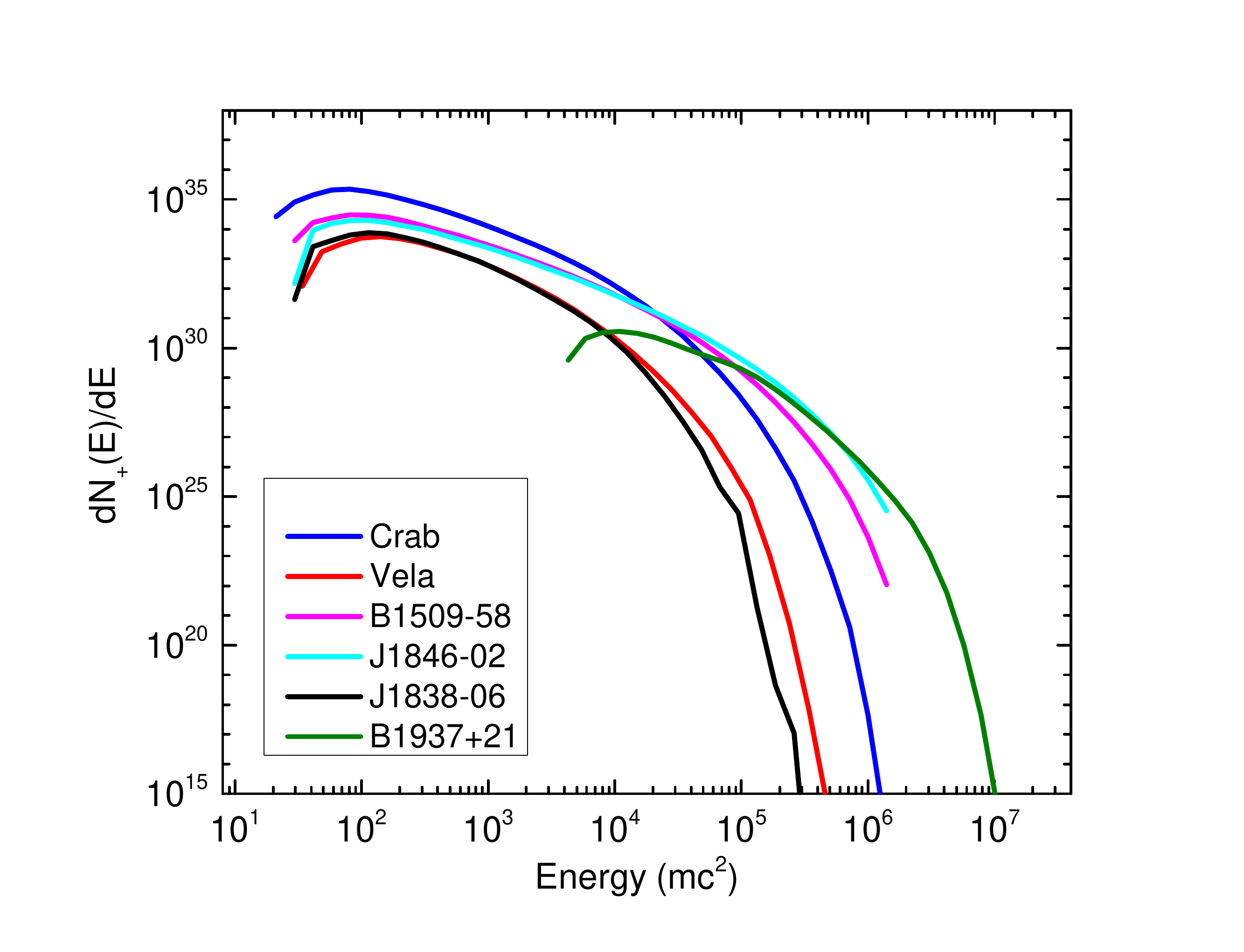}}
     \caption{Electron-position pair spectra, number of pairs per s per energy interval from one polar cap.}
      \label{fig:pairs}
     \end{figure}

Our model for MeV pulsars uses the code developed by \cite{HK2015} to compute synchrotron-self Compton (SSC) spectra for the Crab and Crab-like pulsars.  This calculation follows trajectories of electron-positron pairs from polar cap cascades, injected near the neutron star surface in a near force-free magnetosphere.  The pair spectra are first computed using a local Monte-Carlo pair cascade simulation \cite{HM2011,TH2015}, where primary electrons are accelerated by strong near-surface electric fields.   Their curvature radiation (CR) photons are attenuated by the one-photon pair production process in the strong magnetic field, and the resulting pairs radiate synchrotron emission while losing their momentum perpendicular to the magnetic field.  The cascade terminates after four or five pair generations when photons drop below pair production escape energy, and the resulting pair spectrum that propagates to the outer magnetosphere is the distribution of pair energies after synchrotron losses.  Examples of polar cap pair spectra of several pulsars, including the MeV pulsars we study here, are shown in Figure \ref{fig:pairs}.  The pair spectra extend over several decades in energy, from Lorentz factors of around 20 up to $10^5$ in the case of J1838-0655 and Vela,  up to $10^{6}$ for the Crab, and beyond $10^6$ for B1509-58 and J1846-0258.  In the case of the young pulsars, the maximum energy of the pair spectra is higher for pulsars with shorter periods and higher surface magnetic fields.  Of the pulsars in Figure \ref{fig:pairs}, B1509-58 and J1846-0258 have the highest magnetic fields and also the highest maximum pair energies.  The pair spectra of   energetic millisecond pulsars (MSPs), such as B1937+21 shown in Figure \ref{fig:pairs}, can extend to maximum energies of $10^7$, higher even than for young pulsars.  Their fast rotation rates produce large polar caps, small curvature radii and therefore high CR energies that are necessary to produce the energetic pairs in their low magnetic fields.  These energetic MSPs are also MeV pulsars and are GeV bright.  

The pairs lose their perpendicular energies via SR in the strong magnetic fields very close to the neutron star, but they retain their parallel energies as they propagate into the outer magnetosphere.  There, they can undergo cyclotron resonant absorption of radio photons when the magnetic field has dropped enough to put the radio photons in the cyclotron resonance in their rest frame \cite{LyubarskiPetrova1998,Harding2008}.  This process excites the particles to higher Landau levels, and they can again emit SR.  This process can be modeled for radio-loud pulsars where the radio flux and luminosity are known.  However, for the radio-quiet MeV pulsars where there is no measured radio flux, we have instead modeled the SR from pairs by giving the particles an assumed pitch angle, $\psi$ in the outer magnetosphere between radii of $0.7\,R_{\rm LC}$ and $2.0\,R_{\rm LC}$.  We also model CR from primary electrons accelerating in a constant electric field, ${\cal{E}}_\parallel = eE_\parallel/mc^2$, with trajectories along the last open field line that extends to $2.0\,R_{\rm LC}$ in the current sheet.  The force-free magnetic field values are interpolated from a numerically computed grid.  All the particle trajectories and radiation are followed in the inertial observer's frame, where the magnetic field values are interpolated.

\begin{figure} 
\center{\includegraphics[width=.8\textwidth]{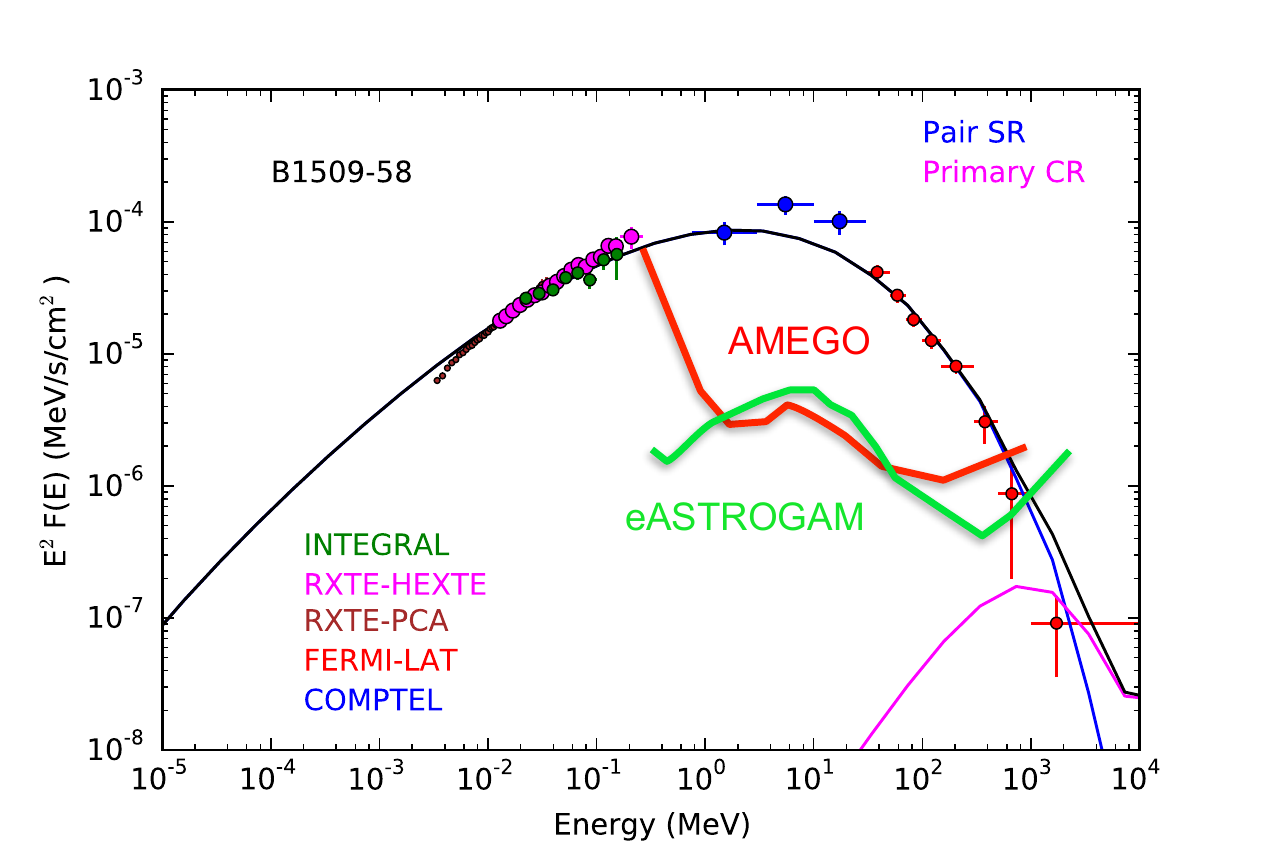}}
     \caption{Model spectral energy distribution for PSR B1509-58 and data from \cite{KuiperHermsen2015,KuiperHermsen2017}.  Also shown are the one-year sensitivities of AMEGO \cite{Moiseev2017} and e-Astrogam \cite{DeAngelis2017}.}
     \label{fig:B1509}
     \end{figure}

\begin{figure}
     \includegraphics[width=.55\textwidth]{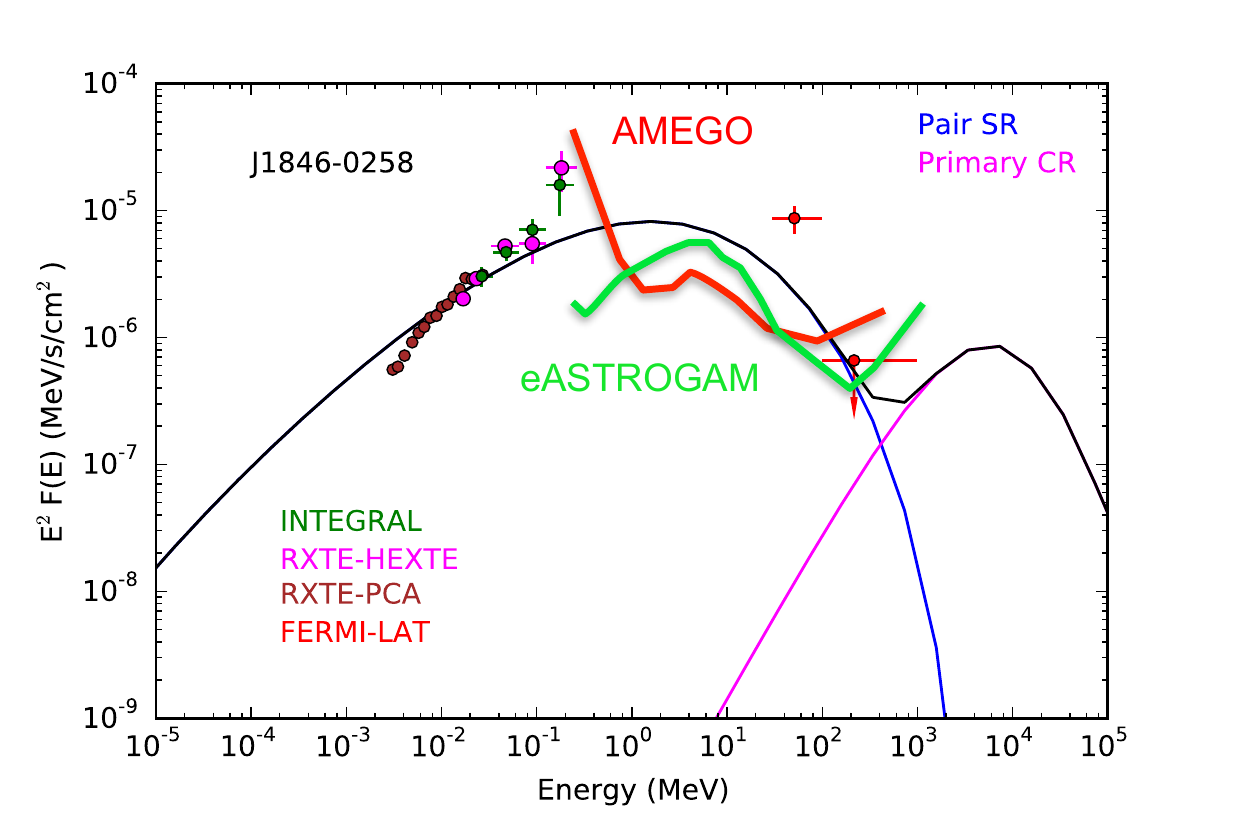}\hspace{-0.6cm}\includegraphics[width=.55\textwidth]{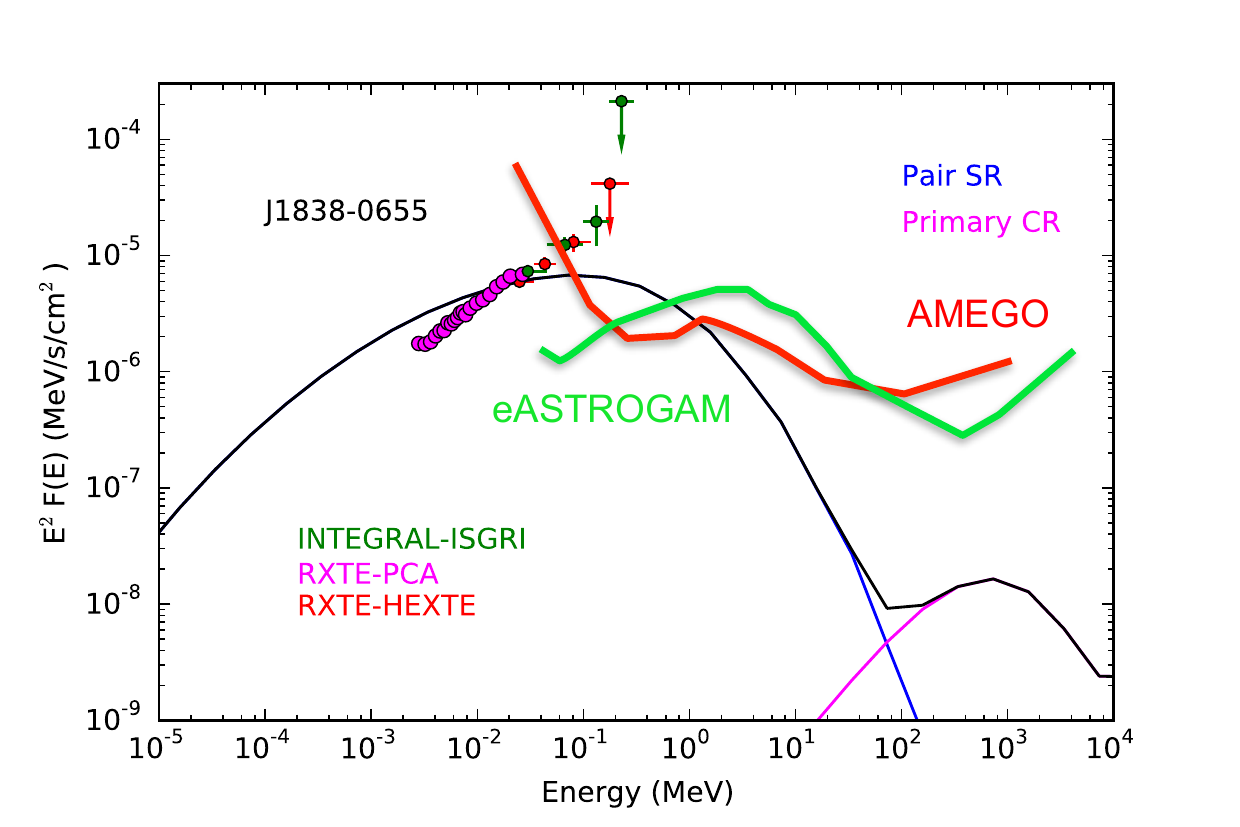}
     \caption{Model spectral energy distribution for PSR J1846-0258 and J1838-0655 and data from \cite{KuiperHermsen2015,KuiperHermsen2017}.   Also shown are the one-year sensitivities of AMEGO \cite{Moiseev2017} and e-Astrogam \cite{DeAngelis2017}.}
      \label{fig:J1846}
     \end{figure}

Figure \ref{fig:B1509} shows our model spectrum from pairs and primaries for PSR B1509-58, along with X-ray and Fermi-LAT data.  In this case an assumed particle pitch angle $\psi =  0.2$, a pair multiplicity of $10^5$ and inclination angle $\alpha = 30^\circ$ produces SR that matches the measured SED, which peaks around 2-3 MeV.  The recent Fermi data \cite{KuiperHermsen2017} severely constrains the CR so that the ${\cal{E}}_\parallel  <  0.1\,\rm cm^{-1}$.  PSR B1509-58 thus appears to be emitting primarily SR from pairs up to an energy of a GeV.  In Figure \ref{fig:J1846} we show model spectra for PSR J1846-0258, for particle pitch angle $\psi =  0.1$, pair multiplicity of $10^5$ and inclination angle $\alpha = 30^\circ$, which peaks around 2-3 MeV and is very similar to the B1509-58 spectrum.  Again, the CR component is somewhat but not as severely constrained by the Fermi-LAT data, so that ${\cal{E}}_\parallel  <  0.5\,\rm cm^{-1}$.  The model spectrum of PSR J1838-0655, for particle pitch angle $\psi =  0.7$, pair multiplicity of $10^5$ and inclination angle $\alpha = 30^\circ$, also shown in Figure \ref{fig:J1846} peaks at a much lower energy, around 100 keV, reflecting the pair spectrum that cuts off at a lower energy that the pair spectra of B1509-58 or J1846-0258.  For this pulsar there is no measured Fermi data and thus, no constraint on the shape of the SR SED or on the CR component.

The SED peak of SR produced by pairs in the outer magnetosphere is expected to be $E_{\rm SR} \sim \gamma_{+}^2\,B_{\rm LC}$, where $\gamma_+$ is the maximum energy of the pairs and $B_{\rm LC} = B_0\,(R_0/R_{\rm LC})^3$ is the magnetic field strength at $R_{\rm LC}$.  
In this model, where the pairs are produced by polar cap cascades and radiate near and beyond the light cylinder, we expect that the SED peak of the SR will be roughly,
$E_{\rm SR} \propto B_0\,B_{\rm LC}$.  In Table 1 we list the values of $B_0$ and $B_{\rm LC}$ for several MeV pulsars.  The $B_0$ values of B1509-58 and J1846-0258 are both very high, but differ by a factor of 3.  However, their values of $B_{\rm LC}$ also differ by a similar factor but the product of the two is nearly the same, predicting roughly the same SED peak energies.  PSR J1838-0655 has a much lower $B_0$ and a $B_{\rm LC}$ comparable to that of B1509-58, thus our model would predict a lower peak of the SR SED.  A model where pairs are produced in the outer gap would have a predicted SED peak of $E_{\rm SR} \propto B_{\rm LC}^{7/2}$, with no dependence on $B_0$.  This model would predict that the SED peak of B1509-58 would be much higher than that of J1846-0238.  With the existing data, this does not seem to be the case.  However, with better spectral measurements at MeV energies, the origin of the emitting pairs could be constrained.

\vskip 0.5cm
\begin{tabular}{ccccc}
& & {\bf Parameters of MeV Pulsars} \\
PSR & P [ms] & $\log(\dot{E})$ [erg/s] & $B_0$[$10^{12}$G] & $B_{LC}$[$10^{4}$G]  \\
\hline
J1838-0655 & 70.5 &  36.75 &  3.80 & 9.76  \\
J1849-0001 &  38.5 &   36.99 &  1.49 &  23.6 \\
J1846-0258 & 324 & 36.91 & 96.5  &  2.55 \\
B1509-58 & 150 &  37.26 &  30.9 &  8.25 \\
\hline
\label{tab:regions}
\end{tabular}
\vskip 0.5cm

Why are a number of MeV pulsars GeV-quiet?  The GeV emission requires particles that are accelerated to high energies, at least $\gamma \sim 10^7$ for CR, and to observe this emission our viewing angle must traverse at least part of the region of acceleration.  Recent models of the global magnetosphere locate this region of acceleration along or near the current sheet outside the light cylinder \cite{PhilippovSpitkovsky2017,Kala2017}.  In the case of young, energetic pulsars, the accelerating layer along the current sheet is very narrow.  A range of viewing angles, particularly for smaller inclination angles will not cut through this narrow layer and thus not observe the GeV emission.  Furthermore, the light curves for small inclination tend to be broad, single peaks rather than narrow double peaks \cite{Kala2014}.  Thus, the properties of the GeV-quiet MeV pulsars may be a consequence of their having small magnetic inclinations and viewing angles that miss both the radio beam and the narrow layer of particle acceleration.
     
\section{Polarization Characteristics of MeV Pulsars}

\begin{figure} 
     \includegraphics[width=.8\textwidth]{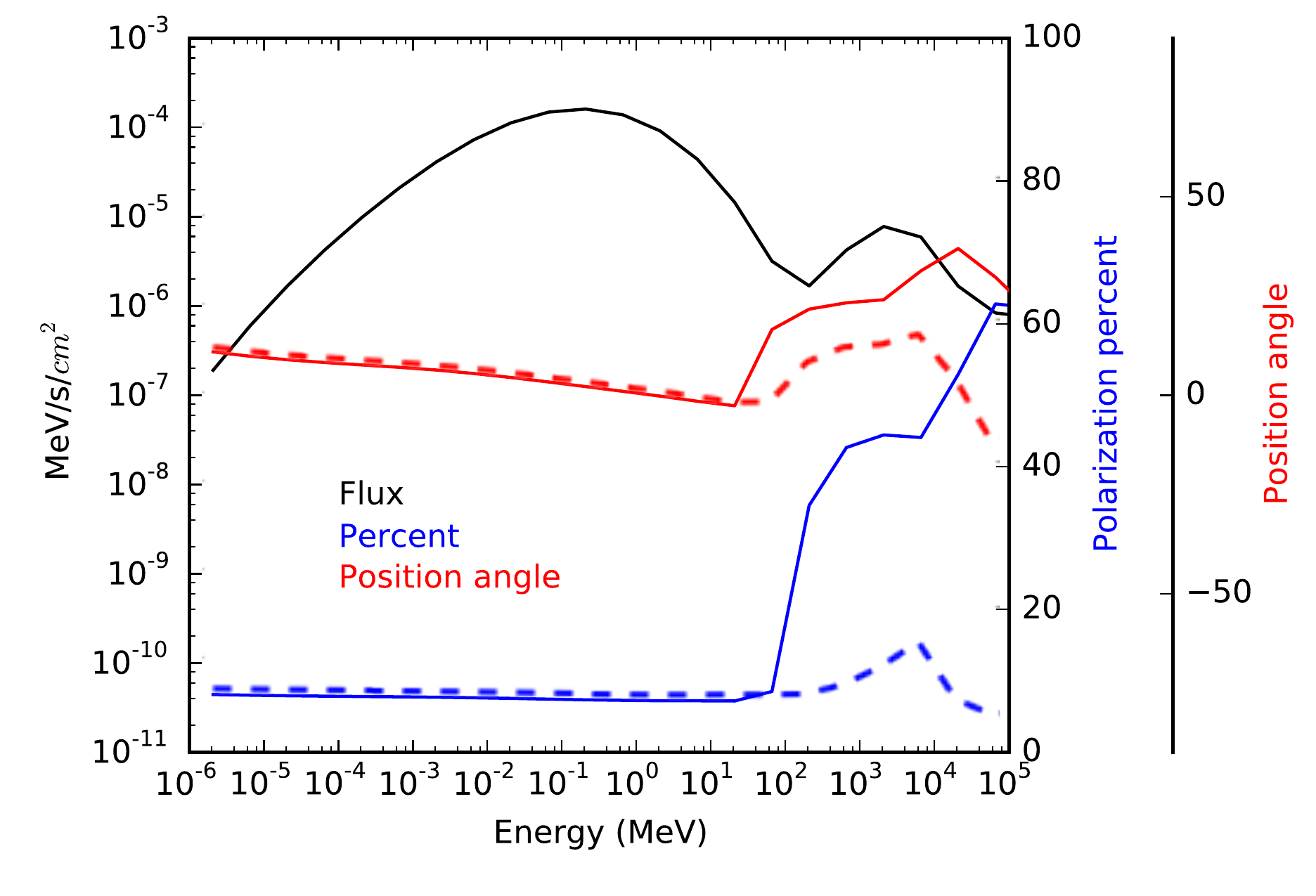}
     \caption{Model spectral energy distribution, position angle and polarization degree for a Crab-like pulsar.}
     \label{fig:CrabPol}
     \end{figure}

Polarization of MeV and GeV emission is a powerful, independent diagnostic, capable of constraining both the location and mechanism of the radiation.  We have modeled the expected polarization characteristics of the emission of MeV pulsars in the model described above.  A detailed description of this calculation is given in \cite{HK2017}.  For Crab-like pulsars where both MeV and GeV emission components are observed, it is possible to detect polarization signatures at the transition between the two components.  Figure \ref{fig:CrabPol} shows an example of the expected phase-averaged polarization of MeV and GeV radiation components for a Crab-like pulsar.  The MeV component is SR from pairs and the GeV component is assumed to be either CR (solid red line) or SR (dashed red line) from accelerated particles.   
The polarization degree of the SR is fairly low, around 10\% or less, while the CR polarization degree is much higher, around 40\%.  The lower level of SR polarization is due to depolarization partly by the particle gyromotion \cite{Takata2007} and partly by the caustic peak formation that blends emission from a range of radii with different field directions (\cite{HK2017}.  The position angle of CR is along the direction of the particle acceleration, along the trajectory radius of curvature.  If the GeV component is CR, then one should observe a rise in polarization as well as a change in position angle.  Therefore, it is possible to distinguish CR from SR as the GeV component emission mechanism.   

\section{Summary}

We have presented results of a model for radiation spectra and polarization of MeV pulsars that show high fluxes of non-thermal emission at hard-X-ray energies.  Our model assumes that the hard X-ray/MeV emission is SR from pairs radiating near the light cylinder that are produced in cascades near the polar cap.  The peak of the SR component SED is proportional to the produce of surface and light cylinder magnetic fields, $B_0\,B_{\rm LC}$, since the maximum energy of the polar cap pair spectra strongly depends on $B_0$.  Measurement of the SED peaks for a number of pulsars can test this model and differentiate it from models where the pairs are produced near the light cylinder, where the SR peaks would have a much stronger dependence on $B_{\rm LC}$ and no dependence on surface field.  Thus, planned telescopes with sensitivity at MeV energies, such as AMEGO\cite{Moiseev2017} and e-ASTROGAM \cite{DeAngelis2017}, could determine the origin of the radiating pairs.

We also showed that the different polarization characteristics of SR and CR provides a means of distinguishing between these processes as the GeV emission mechanism.  Polarimeters sensitive in the 1 - 100 MeV range could detect the sudden rise in polarization degree that would signal a transition from SR to CR.  Even in non-Crab-like pulsars such as Vela, detection of a polarization degree greater than 20\% of the GeV emission would indicate CR as the mechanism.  These measurements can thus distinguish between models of outer magnetosphere emission that advocate CR \cite{Kala2017,Kala2014} or SR \cite{Cerutti 2016,Petri2013} for the GeV component.

\end{document}